\documentclass[trackchanges]{aastex701}

\usepackage[version=4]{mhchem}
\shorttitle{N- vs O- interstellar disparity}

\begin{document}

\title{Interstellar disparity: N- vs O-bearing molecules:\\ A multidisciplinary perspective across interstellar environments}

\author[0000-0003-0167-0746]{Mathilde Bouvier}\thanks{These authors contributed equally to this work.}
\affiliation{Leiden Observatory, Leiden University, PO Box 9513, 2300 RA Leiden, The Netherlands}
\email{bouvier@strw.leidenuniv.nl}
\correspondingauthor{bouvier@strw.leidenuniv.nl, j.enrique.romero@lic.leidenuniv.nl, marta.desimone@eso.org}

\author[0000-0002-2147-7735]{Joan Enrique-Romero}\thanks{These authors contributed equally to this work.}
\affiliation{Leiden Institute of Chemistry, Gorlaeus Laboratories, Leiden University, PO Box 9502, 2300 RA Leiden, The Netherlands}
\email{j.enrique.romero@lic.leidenuniv.nl}

\author[0000-0001-5659-0140]{Marta De Simone}\thanks{These authors contributed equally to this work.}
\affiliation{European Southern Observatory, Karl-Schwarzschild-Strasse 2, D-85748 Garching bei München, Germany.}
\email{marta.desimone@eso.org}

\begin{abstract}

Observations of O- and N-bearing molecules reveal systematic chemical differentiation across a wide range of interstellar environments. This pattern is peculiar and its origin likely reflects a combination of physical conditions, chemical pathways, and observational effects rather than a single mechanism. In this paper, we provide a comprehensive overview of the current understanding of the N-bearing and O-bearing species and their relative behaviour in the interstellar medium, integrating results from observations, laboratory experiments, chemical modelling, and theoretical studies. While inspired by discussions at the Lorentz Center workshop held in Leiden in August 2025, this work goes beyond a summary of the meeting and instead synthesizes the state of the field, highlights key open questions, and outlines future directions.  We identify several key observational, experimental, and theoretical challenges that must be addressed to determine the origin of this chemical differentiation.

\end{abstract}


\section{Introduction} \label{sec:intro}

Astrochemistry sits at the crossroads of astronomy, chemistry, and physics and draws on all three fields to address fundamental questions: How does chemistry unfold in space? How does it shape the cosmic environments where stars and planets are born? What is the ultimate level of chemical complexity reached in planet-forming regions?  Did the building blocks of life originate in space, carried along with the processes of star and planet formation?
By tracing molecules in the interstellar medium (ISM), scientists gain a unique window into the hidden processes that drive the formation of galaxies, stars, and planetary systems and, ultimately, into the conditions that may give rise to life itself.

In the past few decades, the field has been transformed by mm-radio interferometers such as ALMA\footnote{Atacama Large submillimeter/millimeter Array (\url{https://almascience.eso.org})}, IRAM-NOEMA \footnote{Northern Extended Millimetre Array (\url{https://iram-institute.org/observatories/noema/})}, the VLA\footnote{Jansky Very Large Array (\url{https://public.nrao.edu/telescopes/vla/})}, and single-dish telescopes like IRAM–30m and Yebes-40m. Thanks to these facilities, about 350 molecules have been detected in the ISM and in circumstellar shells  \citep[\url{https://molecules-in.space}; CDMS,][]{Muller2005CDMS}. Chemistry in the ISM is exotic and surprising, often producing molecules that cannot exist for more than a fraction of a second on Earth, such as radicals and ions. However, under interstellar conditions, it can lead to the formation of a special subset of molecules, the interstellar complex organic molecules (iCOMs). These species, with backbones made of carbon and hydrogen, often containing heteroatoms such as nitrogen (N) and oxygen (O), are very interesting for their potential prebiotic role. They 
are often described as the ``building blocks'' of larger, life-related molecules, making them a natural focal point in the search for connections between interstellar chemistry and the origin of life on Earth \citep{Herbst_vDishoeck_2009, Ceccarelli2023ASPC}.

\begin{figure*}
    \centering
    \includegraphics[width=0.9\linewidth]{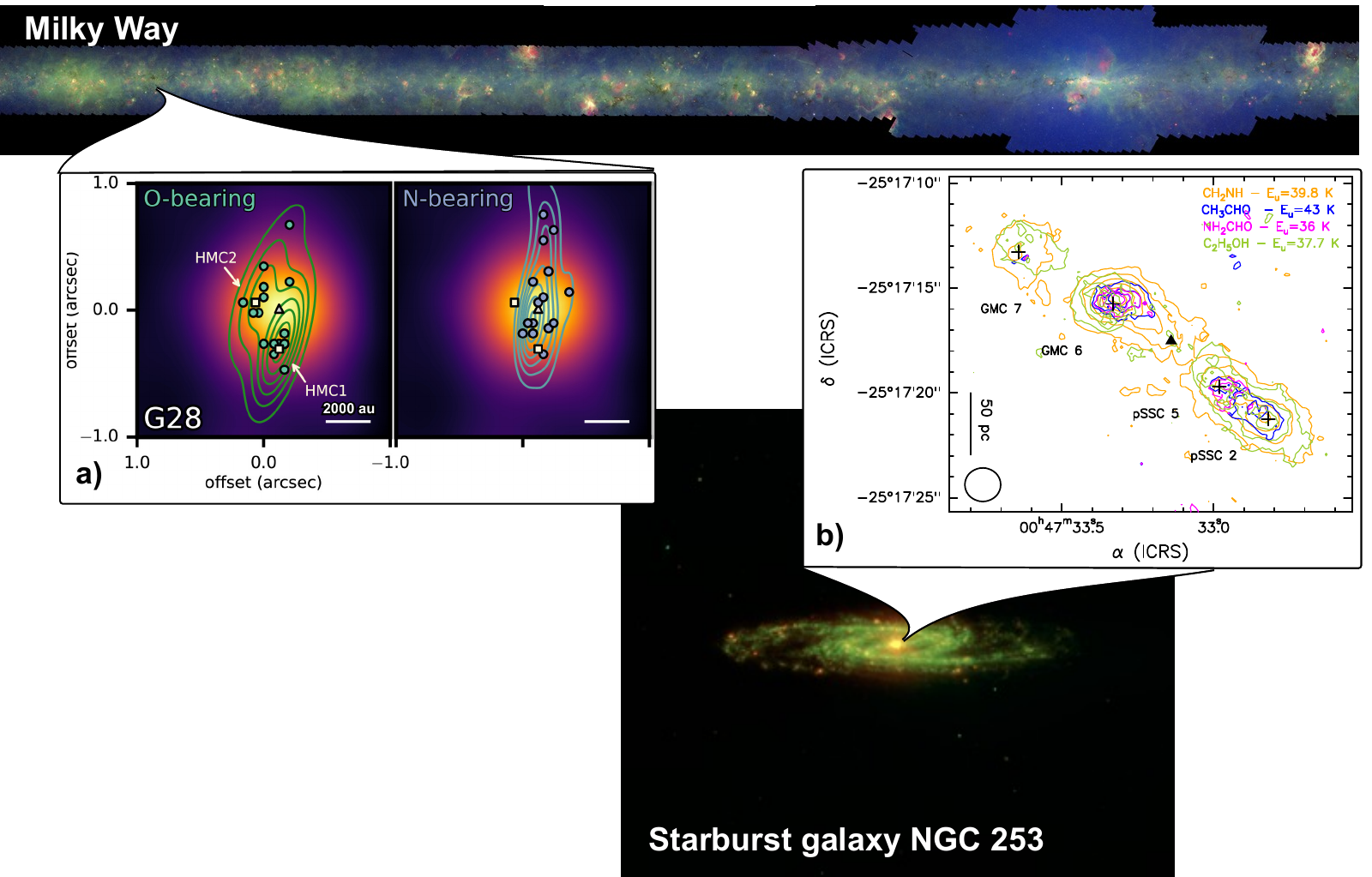}
    \caption{Two examples of O- and N-bearing iCOMs dichotomy. On the top is shown a three-color composite of the Milky Way from Spitzer observations. The 3.6-micron, 8-micron and 24-micron lights are shown in blue, green, and red, respectively. Credit: NASA/JPL-Caltech/Univ. of Wisconsin. (a) Within the MW, the massive star forming region G28.20-0.05 (G28) is a massive star-forming region containing two hot molecular cores (HMC1 and HMC2) showing distinct emission morphology of O- and N-bearing iCOMs. Adapted from \cite{Law2025}. On the bottom is shown the nearby starburst galaxy NGC\,253, in 8-micron and 24-micron lights in green and red respectively, also from Spitzer observations. Credit: NASA/JPL-Caltech. (b) In its central molecular zone, giant molecular clouds where massive star-formation is ongoing, also show different spatial distribution of O- and N-bearing iCOMs. Adapted from \cite{bouvier2025}.}
    \label{fig:dichotomy}
\end{figure*}

However, high-resolution observations reveal a persistent imbalance: in some environments, nitrogen-bearing iCOMs dominate, while in others oxygen-bearing iCOMs are far more abundant. This nitrogen vs. oxygen dichotomy is not confined to a single region or condition; it shows up across a wide range of Galactic environments, from the extremely cold darkness of prestellar cores to the warm, turbulent surroundings of young stars \citep[e.g.][]{blake_1987, jimenez-serra_2016, bonfand_2017,codella_2017, Csengeri2018, tercero_2018, mininni_2023, Busch_2024, Law2025}, but also in extragalactic regions \citep{bouvier2025}. Figure~\ref{fig:dichotomy} shows two examples of spatial dichotomy detected, in a Galactic massive star-forming region, and in the giant molecular clouds of a nearby starburst Galaxy as an example. This differentiation suggests that different underlying chemical or physical processes are at work, yet their exact nature remains elusive. 
Whether the observed differentiation reflects a fundamental chemical divide or instead emerges from the interplay of multiple environmental, physical, and observational effects remains an open question.
This uncertainty is more than a technical detail. Understanding why nitrogen- and oxygen-bearing molecules behave so differently can provide important insights into the broader questions introduced above.
The implications extend to the potential link between interstellar chemistry and the emergence of life, since chemical elements such as carbon, hydrogen, oxygen, and nitrogen form the backbone of biomolecules such as amino acids.

In this context, we convened a Lorentz Center workshop entitled ``Interstellar Disparity: N vs O-bearing molecules", held at Leiden University in August 2025. The meeting brought together experts in observational astronomy, astrochemical modelling, laboratory, computational and theoretical chemistry. This manuscript summarizes the discussions that emerged during the workshop, presents the current state of the field, and outlines a set of proposed actions to help address the apparent dichotomy between N- and O-bearing iCOMs. Further details on the workshop organisation and participant demographics are provided in Appendix \ref{app:workshop}.

\subsection{Current Understanding of the O/N Disparity}\label{sec:topics}

\subsubsection{Observational evidence for the galactic O- versus N-bearing dichotomy 
} \label{sec:StateOfArt_NvsO}

Why do O-bearing and N-bearing molecules often appear spatially differentiated in interstellar environments? Observational evidence for such a dichotomy has emerged over the past decades across a wide range of physical conditions, from cold starless/prestellar cores \citep{jimenez-serra_2016, JimenezSerra2021} to warm protostellar regions \citep[e.g.,][]{blake_1987, bonfand_2017, tercero_2018, mininni_2023}, outflows \citep{codella_2017, Busch_2024}, and disks \citep{oberg_2023}.

In starless/prestellar cores, surveys of complex organic molecules in regions such as Perseus and Taurus show a clear difference between O-bearing and N-bearing species. While the abundances of several O-bearing molecules remain relatively similar across sources, N-bearing species vary by more than an order of magnitude \citep{scibelli_2020, Scibelli2024MNRAS}. Deep centimetre wavelength observations of the starless core TMC-1 have also revealed a rich inventory of carbon-bearing molecules, including cyanopolyynes and aromatic species, as well as O-bearing molecules like C$_2$H$_3$CHO, C$_2$H$_3$OH, HCOOCH$_3$, and CH$_3$OCH$_3$ \citep{mcguire_2020, agundez2021bearing, cernicharo_2021, cernicharo_2024}. Spatially resolved observations have further revealed differences in the distribution of several carbon-chain and aromatic species, including benzonitrile and cyanopolyynes, relative to radicals such as C$_n$H and C$_n$N \citep{cernicharo_2023}. Whether O-bearing species exhibit similarly distinct spatial distributions in TMC-1 remains an open question.

In warm protostellar environments, iCOMs are commonly associated with compact regions known as hot corinos \citep{ceccarelli_2004}. Their occurrence is not uniform across star-forming regions, with detection rates of about 50\% in Perseus (PEACHES, \citealt{Yang2021ApJ}) and about 20\% in Orion (ORANGES, \citealt{bouvier2021, bouvier2022}). This suggests that hot corino chemistry may depend on large-scale environmental conditions.
Among the hot corinos that have been spatially resolved, several sources show molecular differentiation \citep{lee2017,lee2022, Bianchi2022, okoda2022, frediani2025} with some N-bearing species, such as formamide (HCONH$_2$), tracing more compact regions than other O-bearing molecules. 

Chemical differentiation is also observed in low-mass protostellar outflows. Single-dish observations already revealed emission from simple iCOMs such as methanol (CH$_3$OH) and acetaldehyde \citep[HCOCH$_3$, e.g.,][]{holdship2019}, but interferometric imaging has shown that different species trace distinct regions within outflows and shocks \citep{codella_2017, codella2020, desimone2020b, lopezsepulcre20242024,  bouvier2025a, robuschi2026}. At the same time, some species show remarkably tight correlations across very different environments. A notable example is the strong correlation between CH$_3$OH and CH$_3$CN, observed from prestellar cores to hot corinos and in outflows and shocks \citep{belloche2020, Yang2021ApJ, chahine2022}, suggesting that chemical links can persist across evolutionary stages despite the apparent dichotomy.

In protoplanetary disks, the detection of iCOMs remains challenging due to the small size of their snowlines. During accretion outbursts, snowlines can move outward, enabling detections of some complex molecules \citep{lee2019}. Current observations indicate that the N-bearing species CH$_3$CN is very compact ($<$100au) and located closer to the disk midplane \citep{ilee2021}, and that its abundance ratio with respect to methanol is lower than those measured for O-bearing species such as \ce{CH_3OCH_3 and CH_3OCHO} \citep{brunken2022}.

Finally, high-mass star-forming regions provide some of the clearest observational evidence of this differentiation. Historically, observations toward Orion KL established the concept of the O- versus N-bearing dichotomy, with different molecular species peaking toward distinct physical components, including the compact ridge and the hot core, reflecting the strong spatial complexity of the region \citep[][]{Peng_2013, tercero_2018}.  
Similarly, high angular resolution studies showed that O-bearing species tend to trace the inner envelope and accretion-related structures, while N-bearing species are often found closer to the protostar and along the outflow direction \citep[e.g.,][]{csengeri2019, jorgensen2020}.
Systematic surveys, such as ATOMS, further show that this type of differentiation is common, although not uniform across the population \citep{qiu2020}. However, deviations are also observed, with some species, such as acetone in Orion KL \citep{Peng_2013}, showing spatial distributions that do not follow the general trend, suggesting a dependence on molecular structure and local physical conditions rather than on elemental composition alone.

Overall, these observations indicate that the apparent dichotomy between O- and N-bearing molecules is widespread, but manifests itself differently across environments and spatial scales. Whether this reflects fundamental chemical differences, environmental effects, or observational limitations remains an open question, and is discussed in the following sections.

\subsubsection{ 
Environmental effects and the extragalactic perspective}\label{sec:role_env}

External galaxies probe a wide range of physical conditions that can differ significantly from those found in our own Galaxy, and be more extreme. These differences can include higher cosmic-ray ionisation rates, higher star-formation rates, or the presence of an active galactic nuclei (AGN), to name a few. Hence, studying iCOMs in these extreme environments provides an opportunity to test whether the chemical trends observed in Galactic star forming regions are still valid in more extreme conditions and to assess how strongly environmental factors shape interstellar chemical complexity. 

Several iCOMs have been detected in various types of galaxies. Towards the Small and Large Magellanic Clouds (SMC and LMC), low-metallicity galaxies, their close distance ($\sim 50-60$ kpc; \citealt{Pietrzynski2013, Graczyk2014}) allows to probe sub-parsec scales, as in our own Galaxy. Several O-bearing iCOMs were detected (e.g., CH$_3$OH, CH$_3$OCH$_3$, CH$_3$CHO, CH$_3$OCHO; \citealt{Heikkila1999, wang2009, sewilo2018, shimonishi2023}), with their emission mostly linked to hot core emission. iCOMs abundances, once scaled to account for the lower metallicity in the LMC and SMC, were found to be comparable to those found in Galactic regions, indicating of a similar iCOMs chemistry at play. A recent comparison of O- and N-bearing iCOM column densities relative to CH$_3$OH across Galactic sources and the LMC/SMC further supports this picture, showing that the molecular emission in the Magellanic Clouds generally follow the same trends observed in the Milky Way despite the lower metallicity \citep{aikawa2026chemistry}. Concerning N-bearing iCOMs, the only clear detection in the LMC is for CH$_3$CN, while NH$_2$CHO is only tentatively detected \citep{sewilo2022, golshan2024, broadmeadow2025}.

In other galaxies, where the typical probed scales are tens to hundreds of pc, several O-bearing iCOMs including CH$_3$OH, CH$_3$CHO, C$_2$H$_5$OH and CH$_3$OCH$_3$ have been detected in starburst galaxies, AGN-starburst composite galaxies, (ultra)luminous infrared galaxies (U-LIRGs) or the lensed quasar PKS1830-211 at z=0.89 \citep[e.g.][]{henkel1987, martin2006, muller2011, aladro2015, saito2017, muller2021, eibensteiner2022, huang2024}. The emission of these iCOMs is usually associated with shocks \citep[e.g.][]{meier2012, martin2011, watanabe2014, eibensteiner2022, huang2024, bouvier2025}. N-bearing iCOMs such as CH$_3$CN, CH$_3$NH$_2$, NH$_2$CHO, HC$_5$N and C$_2$H$_3$CN  were detected mainly towards the starburst galaxy NGC\,253, the lensed quasars PKS1830-211 or the LIRG NGC 4418 \citep{muller2011, aladro2015, costagliola2015, tercero2020, martin2021} and were found to be likely associated with either shocks and/or enhanced UV/X-ray/cosmic ray fluxes in the surrounding medium \citep[e.g.][]{tercero2020, gorski2021, bouvier2025}. Only CH$_3$CN was also detected towards several types of galaxies \citep[e.g.][]{mauersberger1991, martin2011, aladro2011, muller2011, qiu2020, sato2022}, and is typically associated with warm dense gas, similarly to galactic hot cores \citep{martin2011, muller2011, sato2022}. 

In the extragalactic context, it is also worth mentioning smaller species, precursors of iCOMs but considered as relatively complex, such as methanimine (CH$_2$NH) and cyanoacetylene (HC$_3$N).
These two N-bearing species present unusually bright emission or high abundances towards compact obscured nuclei\footnote{compact (r$<$100 pc), hot (T$>$100 K), and opaque ($N_{H2}>10^{24}$ cm$^{-2}$) nuclei in some of (U-)LIRGs  \citep[e.g.][]{sakamoto2010, aalto2015b, falstad2021}} (CONs; e.g. \citealt{aalto2002, aalto2007, aalto2012a,  lindberg2011a, costagliola2011, gorski2023}), environments presenting warm shielded gas from UV and particle radiation.

Finally, whether the N- vs O-bearing dichotomy is seen in extragalactic environments has been addressed by one only study so far: Towards the central molecular zone (CMZ) of NGC\,253, \cite{bouvier2025} found some chemical differences occurring at scales of giant molecular clouds (GMCs; $\sim 30$ pc), with a clear dichotomy between the various iCOMs emission in some of the GMCs, but this dichotomy is likely due to a difference in GMC properties rather than the extreme (high cosmic-ray ionization rates) conditions of NGC\,253.
Overall, when comparing similar scales between Galactic and extragalactic environments, there does not seem to be major differences or surprises regarding the emission of iCOMs, indicating a minimal effect from the extreme conditions. However, the extragalactic studies of iCOMs are still quite limited. The presence of a N- vs O-bearing dichotomy has been found in one study, but the cause of this dichotomy and whether the environment plays a role is not clear and still needs to be investigated.

\subsubsection{Gas–grain chemistry and functional group selectivity}

Traditionally, non-energetic chemistry in the ISM has been described within two main paradigms: gas-phase chemistry and grain-surface chemistry. Early astrochemical models placed greater emphasis on the former. However, both chemical models and laboratory experiments soon highlighted the importance of grain-surface chemistry for the formation of hydrogenated species and also iCOMs \citep[e.g.,][]{charnley1992molecular, watanabe2002efficient, watanabe2003dependence}. This view was further reinforced when it became clear that gas-phase reactivity alone could not efficiently account for the observed abundances of several key species, such as CH$_3$OH \citep[e.g.,][]{Geppert2006}, leading to renewed attention on the role of icy grain chemistry.

These two regimes operate under markedly different physical conditions. Although both occur at very low temperatures ($\sim$10~K), gas-phase astrochemistry occurs at densities of $\sim 10^{4}$~cm$^{-3}$, where three-body collisions are extremely rare and radical–molecule and ion–molecule reactions dominate \citep[e.g.,][]{Herbst_vDishoeck_2009,heays2017photodissociation,balucani2015formation}. In contrast, grain-surface chemistry proceeds on interstellar dust grains, where gas-phase species freeze out onto cold surfaces. Once accreted, atoms and molecules can undergo efficient H-addition reactions, leading to the formation of hydrogenated species such as \ce{H2O} \citep[e.g.,][]{tielens1982model, Dulieu2010, oba2012water} and methanol \citep[e.g.,][]{watanabe2002efficient, fuchs2009hydrogenation, linnartz2015atom}. As gas-phase species accrete onto dust grains and undergo hydrogenation reactions, the products accumulate to form icy mantles. Their composition evolves with cloud evolution: water-rich ices form first, whereas methanol-rich ices are expected to develop later, when CO efficiently freezes out onto grain surfaces in colder and denser environments \citep[e.g.,][]{cuppen2009microscopic,Cuppen2024_review}. Recent JWST observations have also suggested that methanol may form through additional pathways, such as \ce{CH_3 + OH} reactions in less shielded, CO-poor environments, indicating that the dominant formation route may depend on the physical conditions \citep[][]{mcclure2023ice}.

As a consequence, icy mantles act as chemical reservoirs in which reactants are concentrated, while also serving as catalytic media and effective third bodies that enable reactions that would be inefficient in the gas phase. In this way, they contribute significantly to the chemical evolution of star-forming regions. The products formed in the ice can later be released back into the gas phase through thermal or non-thermal desorption processes \citep[e.g.,][]{Dulieu2010,Minissale2022,Cuppen2024_review}.

Despite the central role of hydrogenation reactions on interstellar icy grains, these reaction networks are not strictly one-way processes. Competing H-addition and H-abstraction steps can generate chemical loops that recycle reactive intermediates \citep[e.g.,][]{Minissale2016b,Chuang2016, Oba2018, Nguyen2021}, while side reactions can divert the chemistry toward alternative products. One example is the formation of formamide during the co-hydrogenation of NO and \ce{H2CO} on icy grains \citep{dulieu2019efficient}. Within this framework, the surface hydrogenation of NO is particularly relevant because it links oxygen- and nitrogen-bearing chemistry on icy grains and can produce species such as hydroxylamine \citep[e.g.,][]{congiu2012efficient,minissale2014solid,ioppolo2014solid,nguyen2019experimental, molpeceres2023processing}. This immediately raises a broader question: \textit{where do the oxygen and nitrogen reaction networks intersect and mix? Does this occur primarily in the gas phase, on icy grains, or only with the aid of energetic processing such as cosmic rays?}

A central mechanism proposed for the formation of iCOMs under non-energetic conditions involves the coupling of radical species through Langmuir--Hinshelwood-type surface reactions \citep{Hasegawa1993,Garrod2006,Cuppen2017,Garrod2013}. In this scenario, radical species formed previously through either energetic processing of simple hydrogenated species \citep[e.g., via ultraviolet irradiation][]{Garrod2006} or partial hydrogenation \citep{taquet2012multilayer} diffuse across the ice surface, where they encounter one another and react. However, the detection of iCOMs in very cold regions, where surface diffusion, a key requirement for this mechanism, is expected to be inefficient, challenged this picture \citep[e.g.,][]{Marcelino2007,Oberg2010,Bacmann2012,Bacmann2016}.
As a result, our understanding of interstellar chemistry has evolved toward a hybrid framework. In this view, neither gas-phase nor grain-surface chemistry alone dominates. Instead, multiple processes operate simultaneously, including additional gas-phase routes, radical–ice interactions, energetic processing such as photolysis, radiolysis, and particle irradiation, Eley–Rideal and other non-diffusive surface mechanisms, and other processes that bridge traditional gas–grain chemistry  \citep{Herbst_vDishoeck_2009,Ceccarelli2017,Ceccarelli2023ASPC,Jin2020,ruaud2015modelling, Vasyunin2017ApJ...842...33V, shingledecker2018general,molpeceres2021carbon,Borshcheva2025ApJ...990..163B}. Energetic processing can profoundly affect interstellar ices, both structurally, by inducing amorphization, compaction, and porosity loss, and chemically, by opening reaction pathways in otherwise inert ice components \citep[e.g.,][]{leto2003ly,palumbo2006formation,dartois2013swift,urso2022ion,fedoseev2018cosmic}. Nevertheless, grain-surface physics remains central, and key parameters such as binding energies, diffusion barriers, and activation energies are fundamental to describing chemical evolution.

In this context, an important question is whether the observed differentiation between O- and N-bearing species reflects the distinct chemical behaviour of specific functional groups. Central to this discussion are carboxylic acids (\ce{-COOH}) and nitriles (\ce{-CN}), two chemically relevant functional groups that occupy contrasting positions in interstellar chemistry: the former remains surprisingly scarce despite the richness of O-bearing iCOMs, whereas the latter is widespread and central to the inventory and evolution of N-bearing molecules. This focus is not unique: other functional-group contrasts also raise related questions, such as aldehydes versus ketones, the scarcity of amines compared with cyano-bearing species, and the apparent absence of N–N bonded molecules.

\paragraph{\bf O-bearing species: the carboxyl (-COOH) group.}

Despite the richness of O-bearing iCOMs detected in the ISM and their importance for understanding interstellar chemistry, a striking chemical feature is the scarcity of molecules containing the carboxylic acid functional group (\ce{-COOH}). This is noteworthy from a prebiotic perspective, since carboxylic acids contain the same functional group found in biologically relevant molecules such as amino acids. To date, only a few representatives of this family have been detected, namely formic acid, acetic acid, and carbonic acid \citep[\ce{HCOOH}, \ce{CH3COOH}, and \ce{HOCOOH}, e.g.,][]{sanz2023discovery}, 
in contrast with the rich inventory of interstellar aldehydes \citep[including \ce{CH_3CHO} and \ce{HCOCH_2OH}; e.g.,][]{beltran_2009ApJ...690L..93B, desimone_2017, scibelli_2020}. This contrast is particularly intriguing in light of meteorite studies, where carboxylic acids are found to be more abundant than carbonyl compounds \citep[aldehydes and ketones, e.g.,][]{aponte2020M&PS...55.2422A}, and it may provide useful clues to the chemical pathways connecting interstellar and planetary chemistry.
More generally, the scarcity of carboxylic acids is also remarkable compared with other O-bearing species. For example, methanol is abundant and can serve as a precursors to more complex compounds, including methyl formate and dimethyl ether. The chemistry of interstellar carboxylic acids seems not to follow the methanol-derived pathways. Instead, -COOH-bearing species appear to be more closely related to \ce{CO2}, given their structural similarity and the high abundance of \ce{CO2} in interstellar ices \citep{Boogert2015,mcclure2023ice, molpeceres2025hydrogenation,ishibashi2024proposed,molpeceres2025formic}. This naturally raises a key chemical question: \textit{Why does a highly stable and relevant functional group on Earth remain marginal in interstellar chemistry?}
This likely stems from two factors: first, O-bearing surface chemistry in the ISM is dominated by hydrogenation and other simple atom-addition reactions, such as O-addition; and second, although \ce{-COOH}-bearing species may form on interstellar ice surfaces, for example through the HOCO intermediate \citep{molpeceres2023cracking}, their survival can be limited by competing H-abstraction pathways that lead instead to \ce{CO2} \citep{ishibashi2024proposed, molpeceres2025hydrogenation}.

\paragraph{\bf N-bearing species: the Cyanide (-CN) group.}

Nitriles, in contrast to carboxylic acids, are widely observed in the ISM across many environments throughout the star formation process \citep[e.g.,][]{leGal2014interstellar,vastel2019isocyanogen,long2021exploring,nazari2021complex}, despite the elemental abundance of nitrogen being roughly an order of magnitude lower than that of oxygen \citep{asplund2021chemical}. Small nitriles such as HCN are abundant in the ISM and can act as reservoirs of carbon and nitrogen thanks to the strength of the \ce{C#N} bond. In addition, they are considered key parent species for the formation of larger N-bearing molecules. For example, successive hydrogenation of HCN on interstellar icy grains has been proposed as a route toward methanimine (\ce{CH2NH}) and methylamine (\ce{CH3NH2}), species that are themselves important intermediates in nitrogen chemistry \citep{Theule2011,molpeceres2024carbon,enrique2026methanimine}. The growing inventory of interstellar nitriles now extends beyond simple mono-nitriles to species containing multiple \ce{-CN} groups, including isocyanogen, protonated cyanogen, dicyanopolyynes, and, more recently, malononitrile (\ce{CH2(CN)2}) and maleonitrile (\ce{C2H2(CN)2}) \citep{cabezas2020rotational,marchione2022unsaturated,agundez2023discovery,agundez2024rich}.
However, nitriles are not the only form in which nitrogen can be stored during star formation. While \ce{CN}-bearing species are characteristic and often abundant in the gas phase, progression along the hydrogenation sequence from HCN toward more saturated species such as methanimine and methylamine appears observationally more restricted. In parallel, part of the nitrogen inventory may be retained in icy mantles as ammonium salts, such as \ce{NH4^+OCN^-}, whose \ce{OCN^-} component is commonly associated with CO-rich, dust-processed ice environments. Because these salts desorb only at relatively high temperatures, they may affect interstellar chemistry by controlling when specific N-bearing functionalities become available into the gas phase during different evolutionary stages \citep[e.g.,][]{Noble2013,theule2013thermal,kruczkiewicz2021ammonia,vitorino2024sulphur}.\\

More generally, the abundances of O- and N-bearing iCOMs are also shaped by the dominant reservoirs and sinks of each element. Oxygen is efficiently locked into \ce{H2O}-rich ice mantles, where water is chemically stable. Nitrogen, by contrast, can be stored in several chemically distinct forms. A large fraction may reside in \ce{N2}, which is highly stable and chemically inert, while reduced nitrogen in \ce{NH3} remains more chemically accessible, e.g., it can participate in acid--base chemistry to form \ce{NH4+}-bearing salts. This difference is mirrored in the molecular reservoirs themselves: \ce{N2} is a relatively poor reactive feedstock, whereas oxygen-bearing species such as \ce{O2} can be more readily transformed by hydrogenation and oxidation chemistry. These chemical differences not only affect how much O or N is available, but also which functional groups are available in the gas phase or retained in the ice during star formation.

Finally, nitriles have long been proposed as important precursors in prebiotic chemistry. Cyanide-bearing molecules can participate in reaction networks leading to biologically relevant compounds, including peptidic bonds (and therefore amino acids), nitrogenous bases, and amines \citep{powner2009synthesis, 
becker2019unified, 
rivilla2020prebiotic, 
jimenez2020toward, 
Theule2011, 
raaphorst2025toward, 
sandstrom2023crossroads}. 
In this sense, cyanides may represent a chemically privileged entry point into more saturated and potentially prebiotic molecules. \textit{Understanding the chemical fate of nitriles in the ISM, and the conditions under which they can be converted into more complex N-bearing species, therefore, remains an important open question in astrochemistry.}
Recent studies have revised the gas-phase formation networks of key nitriles such as \ce{CH$_3$CN} and \ce{HC$_5$N}, highlighting that their chemistry is more intricate than previously thought and still requires coordinated observational, experimental, theoretical, and modelling efforts \citep{Giani2023MNRAS,giani2025}. In addition, recent theoretical studies of reactions involving electronically excited N atoms with nitriles and cyanopolyynes further illustrate that the post-formation chemistry of these species can be non-trivial, with possible transformation pathways for species such as \ce{CH$_3$CN} and \ce{HC$_3$N} \citep{mancini2023computational,liang2023reactions}.

\section{ Current Challenges and Open Questions}\label{sec:addQ}
\subsection{Galactic observations }

The observational trends outlined in \S \ref{sec:StateOfArt_NvsO} raise fundamental questions about the nature of the apparent dichotomy between O- and N- bearing molecules. Although spatial differentiation is observed in many environments, it is still unclear whether this behaviour reflects a true chemical dichotomy or instead results from a combination of physical structure, excitation effects, and observational biases.

One important issue concerns dust opacity. In compact and deeply embedded sources such as hot corinos, millimetre continuum opacity can partially obscure molecular emission \citep{desimone_2020ApJ...896L...3D} . Since different molecules probe different temperature layers and spatial regions \citep{frediani2025}, line emission originating close to the protostellar center may be more strongly attenuated than emission arising farther out. Consequently, the apparent compactness or absence of some species may not necessarily reflect intrinsic chemistry alone.
Thermal structure and sublimation also likely play a major role. Molecules with different binding energies desorb at different temperatures and therefore trace different regions within protostellar envelopes and disks. In sources with strong temperature gradients, this naturally produces spatial differentiation even between molecules with related formation pathways. In this context, accurate determinations of binding energies and diffusion barriers are essential for interpreting molecular distributions \citep{Ferrero2020, tinacci2023theoretical, kakkar2025ApJ...993..184K}. 
Another open question concerns the relative contribution of grain-surface and gas-phase chemistry. Some iCOMs are thought to form mainly on icy grains and to be later released into the gas phase, while others may form efficiently through gas-phase reactions after desorption of simpler precursors. In shocks and outflows, where grain mantles are sputtered and gas-phase chemistry becomes enhanced, chemical differentiation may therefore reflect differences in chemical timescales, destruction pathways, or shock evolution \citep{codella_2017, desimone2020b, bouvier2025a}. 
At the same time, some observational trends challenge the idea of a strict separation between O-bearing and N-bearing chemistry. A notable example is the strong correlation observed between CH$_3$OH and CH$_3$CN across a wide range of environments. Recent revisions of the chemical network of CH$_3$CN, supported by quantum chemistry calculations, have revealed either a direct chemical link between these two species \citep[][]{Giani2023MNRAS} or a similar chemical origin \citep[][]{enrique2024complex,ER+TL2025}. These results illustrate that correlated abundances do not necessarily imply identical origins, but may reflect a less straightforward chemical link.

Environmental effects further complicate the picture. The different detection rates of hot corinos in regions such as Perseus and Orion \citep{Yang2021ApJ, bouvier2022} suggest that large-scale conditions, such as radiation field, density structure, and dynamical history, influence the emergence and observability of complex organic chemistry. Disentangling environmental effects from evolutionary trends remains challenging and requires systematic and homogeneous observational approaches.

Taken together, these considerations suggest that the observed differentiation between O-bearing and N-bearing species likely results from multiple interconnected effects rather than from a single mechanism. This discussion motivated several central questions during the workshop:
\begin{itemize}
    \item[-] The spatial distribution of O-bearing and N-bearing species often differs across environments and spatial scales. \textit{Does this reflect a true chemical dichotomy, or is it driven by physical conditions, binding energies, or observational biases?}
    
    \item[-] The emission properties of iCOMs vary with environment and evolutionary stage. \textit{Are these variations mainly set by local physical conditions, formation pathways, or both?}
    
    \item[-] \textit{What observations, laboratory data, quantum calculations, and modelling efforts are still needed to distinguish between these scenarios?}
\end{itemize}

\subsection{Extragalactic observations }

The overview presented in \S~\ref{sec:role_env} of the detection of iCOMs and their key precursors indicates that (1) studies in extragalactic environments are very limited to a few galaxies and a handful of iCOMs and (2) the extreme conditions in external galaxies do not seem to drastically alter the physical conditions probed by the iCOMs. In extragalactic environments, chemistry  is primarily used as a diagnostic tool to probe physical conditions. One cannot really use the physical conditions (usually unknown) to study the chemistry, contrarily to Galactic studies.
Hence studying iCOMs and their precursors in extragalactic environments and using what is known about their chemistry, could help addressing several fundamental questions about galaxy formation and evolution.

The detection of a high abundance of methanimine (CH$_2$NH) towards the galaxy IC 860 \citep{gorski2023} as well as the detection of CH$_2$NH maser towards a sample similarly compact deeply embedded Nuclei \citep[][]{gorski2021} raises the question of whether this species could help unveil what powers these objects. Indeed, a significant part of luminous and (ultra-)luminous infrared galaxies (ULIRGS and LIRGS) host such CONs (compact obscured nuclei; e.g. \citealt{sakamoto2010, aalto2015b, falstad2021}) but their nature remains uncertain \citep{falstad2021}. The nuclei may contain a super massive black hole (SMBH) with a high rate of accretion (AGN) or an abnormal mode of star formation \citep{aalto2015b}. If CONs are powered by an AGN, then they likely represent a phase of rapid accretion onto the SMBH embedded in an obscured material. Understanding how galaxies grow SMBHs and what their relationship is to the host galaxy is a fundamental question in galaxy evolution \citep[e.g.][]{SandersMirabel1996, FerrareseMerritt2000, Fabian2012}.

Outflows driven by AGNs and/or starbursts represent a strong and direct mechanism
for feedback that may clear central regions of fuel for star formation or black hole (BH)
growth. Hence, investigating the physical and chemical conditions of the outflowing molecular gas can help us understand the driving mechanism and the origin of the gas and, ultimately, galaxy evolution \citep{Veilleux2020}. To the best of our knowledge, there are no iCOMs detected in these large-scales outflows. Smaller species, and hence precursors of more complex species are, on the other hand, detected. While some N-bearing species such as CN, HNC and  HCN are enhanced in (often AGN-powered) outflows, O-bearing species such as CO, HCO$^+$ and CH$_3$OH are suppressed or absent in outflows \citep[e.g.][]{sakamoto2014, aalto2015a, Garcia-Burillo2015, Cicone2020, Saito2022a, Saito2022b}. However, statistical studies of galactic molecular outflows are still missing today, as current studies are biased towards highly active or gas-rich galaxies, where outflows are likely brighter and easier to observe. 

Nonetheless, these considerations led to the following additional questions raised during the workshop, which motivated proposed actions summarised in \S ~\ref{sec:prop_actions}.
\begin{itemize}
    \item[-] \textit{What do we expect from iCOMs in obscured AGNs? Can they help us constrain SMBH growth and coevolution, and find hidden growing SMBHs?}
    \item[-] \textit{Do molecular species in extragalactic outflows behave similarly as in Galactic outflows? What drives the apparent excess of N-bearing species over O-bearing species in  extragalactic outflows?}
    \item[-] \textit{Are N-bearing iCOMs more prone to be abundant in high temperature and stronger shocks?}
\end{itemize}

\subsection{Laboratory experiments on iCOMs formation}
Experimental astrochemistry provides a powerful way to test hypothetical processes and uncover new chemical pathways under controlled laboratory conditions. The design of the experimental setup is therefore crucial, as it determines which processes can be investigated (gas-phase or solid-state chemistry), how reactions occurring in the chamber can be detected, and which physical conditions can be simulated. Ultimately, it also defines how closely laboratory experiments can reproduce the environments found in the ISM.

In the gas phase, two main experimental techniques are currently employed: CRESU \citep[Cinétique de Réaction en Écoulement Supersonique Uniforme][]{cresu1,cresu2,cresu3} and crossed molecular beam experiments \citep{CrossedBeams1_lee1987molecular,CrossedBeams2_casavecchia1999crossed}. CRESU experiments reproduce the very low temperatures characteristic of interstellar environments (tens of K) with high accuracy, but they probe reactions under bulk conditions and therefore cannot isolate individual collision events. In contrast, crossed molecular beam experiments simulate single collision events, providing detailed information under well-controlled density conditions, although constraining the effective temperature of the reactive encounter is more challenging. CRESU studies have revolutionized astrochemistry in several ways: first, by demonstrating that neutral-neutral reactions can proceed efficiently at very low temperatures, and second, by revealing strong deviations from Arrhenius behaviour in this regime. In several cases, this enhanced low-temperature reactivity has been attributed to quantum tunnelling through activation barriers, often mediated by the formation of weakly bound pre-reactive complexes \citep[e.g.,][]{shannon2013accelerated,heard2018rapid,gao2018kinetics,balucani2024can}. Crossed molecular beam experiments, on the other hand, have proven exceptionally useful for determining reaction mechanisms and branching ratios. Detection techniques vary depending on the species studied and include laser-induced fluorescence (LIF), resonance-enhanced multiphoton ionization (REMPI), time-of-flight analysis (TOF), and mass spectrometry, among others \citep[e.g.,][]{cooke2019experimental}.

On the other hand, for solid-state astrochemistry, the experimental setup typically consists of a high-vacuum chamber equipped with a cold finger that can reach temperatures of around 10~K, thereby mimicking the conditions in the cold, dense ISM. At the end of the cold finger, there is usually a substrate surface (e.g., gold) onto which interstellar ice analogues are deposited. The chamber is connected to several gas inlets that allow control of the ice composition and deposition either simultaneously or sequentially. For example, one may first deposit CO and subsequently expose the surface to atomic H to study the hydrogenation of CO. Compared with gas-phase experiments, fewer detection techniques are available for solid-state studies. The most commonly used methods are in situ reflection-absorption infrared spectroscopy (RAIRS) and mass spectrometry following thermal desorption. Other techniques exist, although they are less widely used \citep[see e.g.,][]{allodi2013complementary,tsuge2023radical}.

In both gas-phase and solid-state studies, increasing chemical complexity inevitably requires controlling a larger number of experimental parameters, which in turn complicates the interpretation of the results. For this reason, laboratory experiments are usually designed to address specific questions under simplified and well-controlled conditions rather than attempting to reproduce the full complexity of the ISM. 
For instance, in solid-state astrochemistry experiments, ice analogues are often prepared as pure ices or simple mixtures containing only a few molecular species, even though real interstellar ices contain many components. This simplification allows one to isolate specific processes, such as the hydrogenation of CO or the formation of more complex species from well-defined precursors. A similar strategy is adopted in gas-phase experiments. Studies performed in CRESU or crossed molecular beam setups typically focus on individual elementary reactions involving a small number of reactants to determine accurate rate coefficients, reaction mechanisms, or branching ratios. While this approach does not capture the full chemical complexity of astrophysical environments, it provides the fundamental data required to construct and validate astrochemical reaction networks.

\begin{figure}
    \centering
    \includegraphics[width=0.9\linewidth]{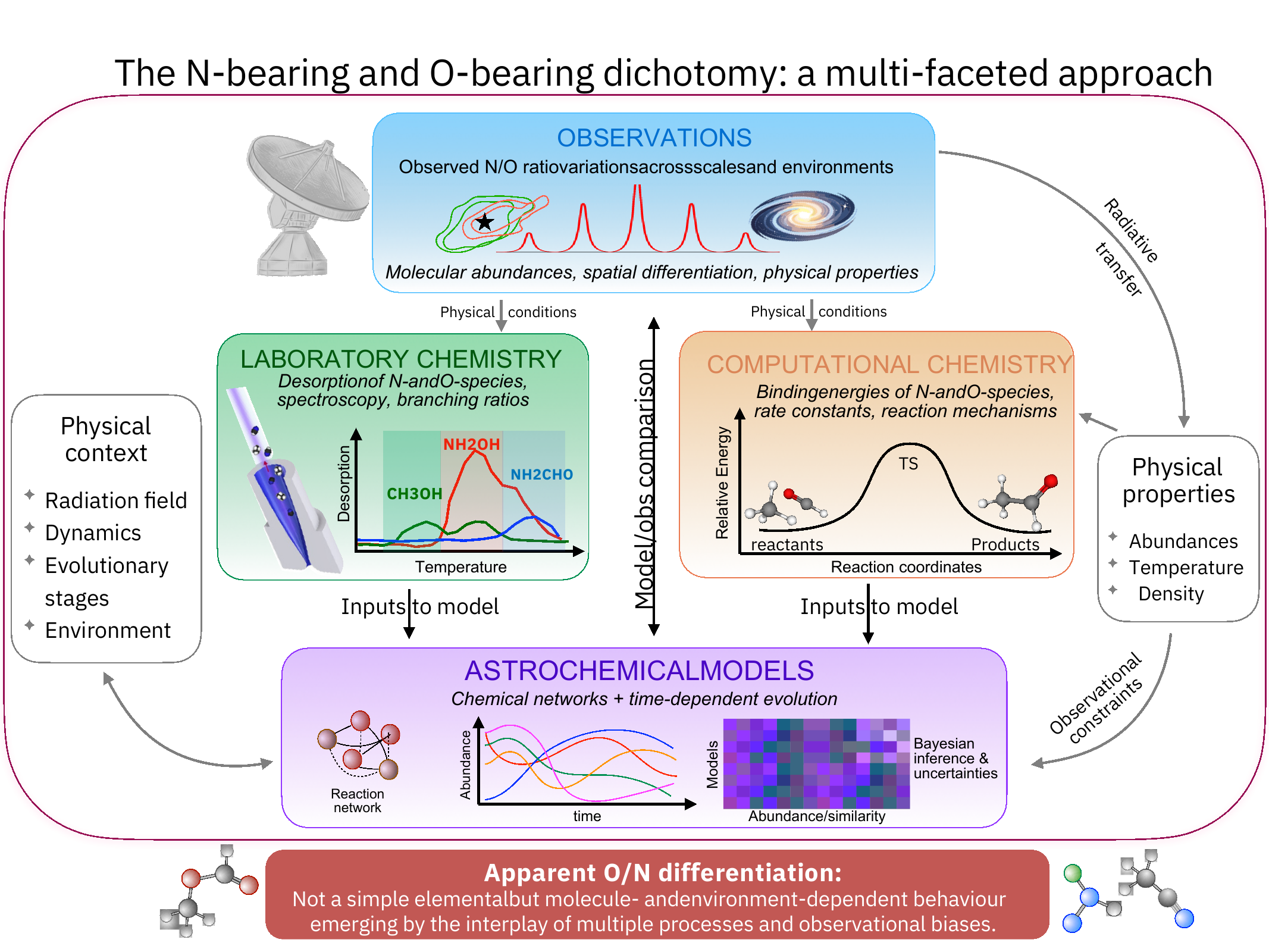}
    \caption{Schematic of the interdisciplinary framework addressing the N-bearing and O-bearing chemical dichotomy.  This diagram illustrates how observations, laboratory experiments, computational quantum chemistry, and astrochemical models mutually constrain one another. By accounting for physical conditions (e.g., radiation fields, dynamics) and observational biases (e.g., radiative transfer, sensitivity limits), this integrated approach reveals the complex physical and chemical feedback processes at the root of the apparent dichotomy.}
    \label{fig:sketch_pillars}
\end{figure}

Experimental astrochemistry, therefore, constitutes one of the three main pillars of the field, together with astronomical observations and theoretical/computational modelling, as shown in Fig.~\ref{fig:sketch_pillars}. The interplay between laboratory studies and these other approaches is particularly powerful. Laboratory data can guide observational searches by identifying plausible formation pathways and spectroscopic signatures of new species. At the same time, theoretical calculations provide mechanistic insight and quantitative parameters that connect laboratory measurements with astrophysical environments. Together, these complementary approaches provide a powerful framework for advancing our understanding of chemical processes in the ISM. A key question then arises: \\
\textit{Can laboratory data guide the search for species, and can observations help solve puzzling laboratory results?}

One area where laboratory measurements can provide crucial quantitative constraints is the determination of binding energies of species on interstellar ice analogues. These quantities determine how long molecules remain on grain surfaces and how easily they can diffuse and react. Binding energies are typically derived from temperature-programmed desorption experiments, often combined with simple kinetic modelling. A clear example is the work of \citet{minissale2016direct}, who constrained the desorption and diffusion energies of O and N atoms on amorphous solid water. Their results showed that O is significantly more strongly bound than N on water ice ($E_{\rm des}\sim 1410$~K for O compared with $\sim720$~K for N), implying that O atoms are retained more efficiently on grains while N atoms remain more mobile and volatile. This naturally raises another question: \\
\textit{Is this fundamental difference in binding energies responsible, at least in part, for the chemical differentiation observed between some O-bearing and N-bearing molecules?}

\subsection{Quantum chemical calculations on iCOMs formation}

In the last few decades, quantum-chemical simulations have become an essential tool in astrochemistry due to their predictive power and their ability to provide atomistic information that is often challenging to obtain from laboratory experiments alone. Clear examples include gas-phase reaction mechanisms and barrier determinations, where theoretical studies have helped interpret experimental results and guide astrochemical models.

In these simulations, the goal is to understand the elementary steps of a reaction at an atomistic level of detail. For example, in the context of surface chemistry, this includes processes such as accretion of species onto a surface, hopping events associated with diffusion, chemical reactions between adsorbates, and desorption back into the gas phase. Quantum chemistry provides a range of tools to investigate these processes, depending on the size and complexity of the system under study \citep[e.g.,][]{Cuppen2017}.
Highly accurate wavefunction-based methods can be used to obtain reliable reaction energies and activation barriers for relatively small systems. However, their computational cost rapidly becomes prohibitive as the system size increases, and therefore these methods are most commonly employed in gas-phase studies or for benchmark purposes. A plethora of literature can be found related to these methods, including introductory books \citep[e.g.,][]{szabo2012modern,cramer2013essentials,jensen2017introduction}. Notice that these methods are still subject to active development \citep[see e.g.,][for a set of reviews]{WFT1_wardzala2026multireference,WFT2_bartlett2024perspective,WFT3_nagy2024state}. Density functional theory \citep[DFT, see][]{DFT3_geerlings2003,DFT1_jones2015}, the workhorse of modern computational chemistry, provides a practical compromise between accuracy and computational efficiency, with some limitations \citep[see e.g.,][]{DFT2_2_verma2020}, but extremely powerful when used appropriately \citep{DFT4_bursch2022}. As a result, it is widely used to study reactions on cluster models of interstellar ice surfaces.
More recently, machine-learning interatomic potentials \citep[MLIP, see e.g.,][]{MLIP1_mishin2021machine,MLIP2_jacobs2025practical}, trained on high-level quantum-chemical data, have emerged as a powerful approach for extending simulations to larger systems and longer timescales while retaining near \textit{ab initio} accuracy. These methods make it possible to explore processes such as diffusion, energy dissipation, and reaction dynamics on realistic models of interstellar ice surfaces \citep[e.g.,][]{molpeceres2020neural,zaverkin2022neural,molpeceres2023reaction,bovolenta2025co,carlos2026reactive}. Their main drawback is the need to train the machine learning model on quantum chemical-quality data, which is still a rather expensive process in computational terms.

Examples of situations where computational chemistry has shown its value include the determination of binding-energy distributions on amorphous ice surfaces \citep[e.g.,][]{Wakelam2017binding,Bovolenta2022,tinacci2023theoretical}, the characterization of reaction mechanisms and activation barriers \citep[e.g.,][]{Zamirri2019,Lamberts2019,zaverkin2022neural,ER2022ApJS,vazart2020gas,puzzarini2026chemistry}, and the calculation of kinetic parameters that can be directly incorporated into astrochemical models \citep[e.g.,][]{molpeceres2021computational}.

In addition, the atomistic perspective provided by computational chemistry has proved particularly valuable for identifying interaction modes between molecules and interstellar ices that are difficult to access experimentally. Examples include the Eley-Rideal reactivity of carbon atoms on water or ammonia ice surfaces, and the ability of CN radicals to form hemibonded complexes with water molecules. These interactions can significantly alter the subsequent chemistry. For instance, chemisorption of carbon atoms on water or ammonia may lead to the formation of species such as formaldehyde or methylamine \citep{molpeceres2021carbon,molpeceres2024carbon}, while CN and CCH radicals \citep{Rimola2018,enrique2024complex,Perrero2022} hemibonded to water may open pathways toward formamide, vinyl alcohol, and ethanol formation. These findings naturally raise several questions:

\begin{itemize}
\item[-] \textit{How general are these unusual adsorption modes? For example, how do cyanopolyynes interact with interstellar ice surfaces, and how does this affect their surface chemistry?}
\item[-] \textit{If Eley–Rideal reactions between atomic carbon and water ice provide an efficient route toward O-bearing molecules, can their importance be tested or constrained observationally?}
\end{itemize}

Finally, translating laboratory measurements and quantum-chemical results into the language of astrochemical models is not always straightforward. Astrochemical models typically require parameters such as reaction rate coefficients, sticking probabilities, diffusion barriers, or chemical desorption efficiencies. However, laboratory experiments often provide averaged quantities over many microscopic processes, while computational studies tend to focus on specific local configurations or individual reaction pathways. Bridging this gap, therefore, requires careful interpretation and, often, the development of simplified parametrizations that capture the essential physics and chemistry in a form suitable for large-scale astrochemical simulations, raising the natural question: \textit{How do different parameters compare between theory, experiments, and modelling?} 

\subsection{Astrochemical modelling} 

Chemical modelling is essential to decipher the information encoded in molecular observations. While radiative transfer modelling provides column densities, abundances, temperatures, and gas densities, it does not by itself reveal the origin, history, or evolutionary stage of the gas, nor the interplay between gas and dust. Chemical models bridge this gap by linking physical conditions to molecular abundances and by incorporating results from laboratory experiments and quantum chemical calculations (see Fig.~\ref{fig:sketch_pillars}). In this way, they allow us to interpret the observed chemical compositions, test formation scenarios, and explore how surface and gas phase processes shape the chemistry over time. For example, molecules often considered shock tracers, such as SiO, HNCO, or CH$_3$OH, can help reconstruct the shock history of a source (e.g., a protostar or a region in a galaxy; \citealt{kelly2017A&A...597A..11K, DeSimone2022, huang2023, huang2024}). However, no tracer is unique and the same species can trace different energetics, gas components, histories of the gas (e.g., C$_2$H; \citealt{cuadrado2015, garciaburillo2017, holdship2021}). 

\begin{figure}[ht]
    \centering
    \includegraphics[width=0.5\linewidth]{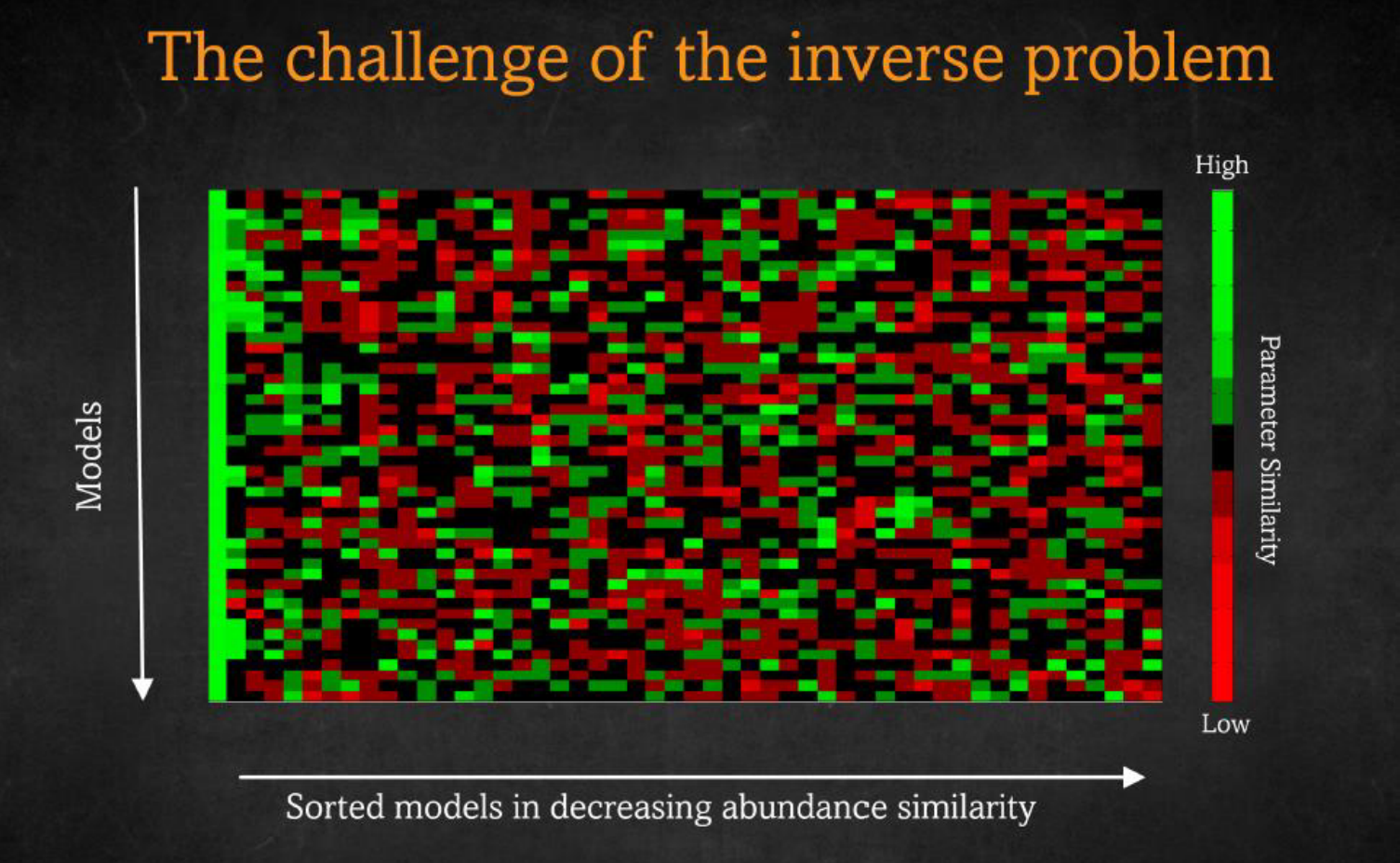}
    \caption{Percentage of abundance similarities for a combination of the most observed species. A hundred thousands time-dependent chemical models varying a large grid of input parameters (e.g. gas density, temperature, cosmic-ray ionization rate, radiation field) were run for $10^6$ yr. The y-axis shows the percentage of similarity and the x-axis sorts the models in decreasing abundance similarity. We can see that many models can lead to very similar output abundances. Credit: Courtesy from S. Viti.}
    \label{fig:inv_problem}
\end{figure}

Chemical simulations are highly sensitive to both physical and chemical parameters. Molecular abundances can vary significantly across the wide range of densities and temperatures present in the ISM. Model predictions depend on several input parameters. For example initial elemental abundances, cosmic ray ionization rates, gas and dust temperatures, reaction networks, desorption mechanisms, and many others. Most of those are still highly uncertain or only partly characterized (like the chemical networks). The large parameter space and the complex interconnections between processes make interpretation challenging and is known as the inverse problem: similar parameters might not give similar abundances and similar abundances might be produced by very different parameters. In other words, solutions may
not be unique or may not depend continuously on the observational data. This problem is represented in Fig.~\ref{fig:inv_problem}, which show how similar the computed chemical abundances can be, even with a large number of models run with a large grid of input parameters. We note that the opposite is also true; many similar models can lead to very different chemical abundances.
Sensitivity analyses,  statistical inference methods and/or machine learning and big data approaches are therefore essential tools to identify dominant reactions, reduce networks when possible, and constrain the most relevant parameters.
Over the years, numerous classical sensitivity analyses have explored the impact of varying physical and chemical parameters \citep[e.g.][]{bayet2008, bayet2011, vasyunin2008, Wakelam2010, woods2012, penteado2017, Furuya2022, dutkowska2025, van_de_sande_2026MNRAS.545f2049V}, typically focusing on restricted parameter sets. Another important source of uncertainty arises from the different physical assumptions, chemical networks, and numerical implementations adopted by astrochemical codes. Benchmarking studies that compare multiple codes under the same physical conditions have shown that, although quantitative differences in the predicted abundances can be not negligible, many of the key chemical trends and dominant formation pathways remain robust across models \citep[e.g.,][]{jimenez2025modelling}. Such comparisons are very valuable not only for identifying the origin of discrepancies but also for distinguishing model-dependent results and provide valuable guidance for improving chemical networks and modelling strategies.

Machine learning techniques and neural networks are also becoming an increasingly useful tool in astrochemistry. They can be used to analyse the output of chemical models, emulate computationally expensive chemical simulations, and  predict molecular abundances from existing chemical inventories.
In the context of chemical modelling, Bayesian inference approaches and systematic sensitivity analyses \citep[e.g.][]{linnartz2015, holdship2018, Heyl2020} together with interpretable machine learning techniques (e.g.,\citealt{vermarien2025}), and complex network analysis \citep[e.g.][]{fernandez2023theoretical, fernandez2025strong}, have been used to explore higher-order dependencies of molecular abundances across large grids of physical and chemical parameters. These studies suggest that, with appropriate caveats, reaction networks can in some cases be reduced or divided into sub-networks, and that when considering a reaction chain, it is crucial to apply constraints to the final or penultimate molecule produced in order to obtain robust conclusions. Recent studies also used combined neural networks and Bayesian inference to infer the gas conditions across the central molecular zone of the nearby starburst galaxy NGC 253 CMZ at 50 pc resolution. The molecular abundances and gas parameter values
were similar to those obtained using a chemical modelling code, but with a significant reduction in processing time \citep{Behrens2024, Behrens2026}. Machine learning techniques have also been applied directly to interstellar chemical inventories. Unsupervised approaches based on vector representations of molecules can identify species that are chemically similar to those already detected, while supervised methods can reproduce observed molecular abundances and predict those of yet-undetected species \citep{lee2021machine, scolati2023ApJ...959..108S}. Another approach has been explored  for the automated identification of molecular emission in complex spectra, using synthetic spectra to train neural networks to recognize molecular signatures \citep{Kessler2025A&A...704A.324K}. Although these methods depend on the quality and range of the training data, they offer a promising avenue to analyse the growing volume of molecular observations, identifying previously overlooked species, and testing whether observed patterns in molecular inventories reflect intrinsic chemical trends or observational biases.

Despite these challenges, chemical models remain indispensable for testing hypotheses and guiding new observations. Finally, it is clear that a complete understanding of the chemistry behind each molecule and its dependencies on the density and temperature of the gas is essential.
At the same time, important open questions remain:

\begin{itemize}
    \item[-] \textit{Should chemical models be used at all, considering the many input and uncertain parameters?}
    \item[-] \textit{Are chemistry knowledge and awareness of the model limitations enough to allow a robust interpretation and proper use of the chemical models?}
    \item[-]\textit{How can we keep open source practices and at the same time avoid the blindly misuse of chemical models? }
    
\end{itemize}

\section{Future directions} \label{sec:prop_actions} 
Progress on understanding the apparent dichotomy between O- and N-bearing molecules requires a \textit{coordinated effort across gas- and solid-phase observations, laboratory work, quantum chemical calculations, and modelling}.

\paragraph{\bf Observations:}
A first observational priority is to establish whether the apparent dichotomy between O- and N-bearing molecules is real, systematic, and widespread, or instead source dependent and molecule dependent. Current evidence is based on a limited number of well studied objects and often relies on specific molecular tracers (for example, CH$_3$CN or NH$_2$CHO). Addressing this requires systematic observational campaigns that target multiple sources across different star-forming regions and evolutionary stages, using consistent spectral setups, sensitivity, and angular resolution. Homogeneous datasets are essential to allow robust comparisons and reduce biases introduced by differing observational strategies.

Spatial resolution is a critical factor. In many cases, the physical scales probed by current observations differ significantly between low- and high-mass star-forming regions, as well as between Galactic and extragalactic sources. Higher angular resolution observations (of less than 0$\farcs1$) in low-mass protostars are required to accurately locate O- and N-bearing molecules and to directly compare them with high-mass sources. Combining interferometric data with single-dish or total-power observations is also essential to recover both compact and extended emission and to avoid artificially enhancing or suppressing apparent chemical differentiation. Expanding the number of well-characterized protostellar and galactic-scale outflows is necessary to determine whether chemical differentiation is a general feature or a consequence of specific physical conditions. When comparing Galactic and extragalactic sources, particular care must be taken to match spatial scales and account for obscuration and beam dilution, as these factors strongly influence the apparent chemical differences.

Observational efforts should also place greater emphasis on simpler molecular species that act as chemical building blocks. While iCOMs are important, their interpretation is often hampered by excitation effects, line blending, and sensitivity limits. Focusing on molecules such as H$_2$CO and CH$_3$OH for O-bearing chemistry, and HCN and HNCO for N-bearing chemistry, provides a more direct way to constrain the initial chemical reservoirs and to assess whether differences arise already at early stages. These species are particularly valuable because their chemical networks are well characterized, allowing more reliable interpretation of observed abundances and spatial distributions. In particular, observations of HNCO and related species may offer key insights into the role of nitrogen chemistry in both gas and solid phases.

Complementary observations of the solid phase are equally important. Many O- and N-bearing molecules, as well as their precursors, can be formed, processed and stored in interstellar ices before being released into the gas phase. The unprecedented sensitivity and spectral resolution of JWST in the infrared are now providing direct measurements of the ice composition toward protostars and molecular clouds, offering new constraints on the molecular reservoirs available for subsequent gas-phase chemistry \citep[e.g.,][]{mcclure2023ice}. Combining JWST observations of interstellar ices with millimetre observations of the gas phase will be essential for linking the solid- and gas-phase chemical evolution \citep[e.g.,][]{Chen2024} and for understanding whether the observed O- and N-bearing differentiation is already established in the solid-phase reservoirs or emerges during subsequent gas-phase evolution. This is particularly relevant for species such as \ce{CH_3OH}, \ce{NH_3}, \ce{OCN^-}, and other ice constituents that are thought to play a key role in the formation of more complex O- and N-bearing molecules.

\paragraph{\bf Laboratory experiments \& computations: }
Progress in understanding the chemical differentiation between O- and N-bearing species requires targeted constraints on key processes in both the gas phase and on icy grain surfaces, together with reliable microscopic parameters for astrochemical modelling.

In the gas phase, one major limitation is the poorly constrained activation of nitrogen. The conversion of \ce{N2} into atomic or radical nitrogen directly affects the formation of N-bearing species, and both experimental and theoretical efforts should therefore quantify the efficiency of reactions that either activate \ce{N2} or bypass it through alternative radical-mediated pathways \citep[e.g., protonation followed by electron recombination primarily returns \ce{N2} back,][]{roueff2015isotopic}. Although the formation of nitriles is thought to proceed efficiently through radical--neutral reactions, their subsequent evolution remains unclear. In particular, it is essential to determine whether nitriles can be converted into more saturated species under interstellar conditions or instead remain chemical end-points because of the stability of the \ce{C#N} bond.

On icy grain surfaces, the main uncertainty concerns the asymmetry between oxygen and nitrogen chemistry. Oxygen-bearing species are efficiently processed through hydrogenation networks, leading to stable products such as water, whereas analogous nitrogen networks appear less efficient and remain poorly characterized. Laboratory studies should therefore quantify hydrogenation pathways for N-bearing species, including branching ratios between addition, abstraction, and fragmentation channels. At the same time, the role of binding energies must be clarified. Because O atoms bind more strongly than N atoms to water ice \citep[e.g.,][]{minissale2016direct,Duflot2021_binding}, they may remain longer on grain surfaces and react more efficiently, but it is still unclear whether this difference alone can account for the observed chemical differentiation. In this context, determining binding-energy distributions, rather than only single representative values, is particularly relevant.

Another key issue is the interaction between reactive species and the ice itself. Non-conventional binding modes, such as chemisorption of carbon atoms or hemibonding of CN radicals, may alter reactivity by stabilizing intermediates or redirecting reaction pathways. Establishing how common these interactions are, and how strongly they affect the chemistry of O- and N-bearing species, is therefore essential.

Addressing these questions requires a combined laboratory and theoretical effort. Experimentally, gas-phase studies should prioritize low-temperature rate coefficients and branching ratios for reactions involving key radicals, while surface studies should focus on controlled hydrogenation and co-deposition experiments, together with quantitative measurements of desorption efficiencies, binding energies, and diffusion barriers. Theoretically, accurate potential energy surfaces and kinetic calculations are needed for both gas-phase and surface processes, including activation barriers, tunnelling contributions, adsorption geometries, and binding-energy distributions on amorphous ice. At the same time, laboratory and theoretical studies should investigate how the form in which oxygen and nitrogen are available, whether as atoms, simple molecules (e.g., \ce{CO}, \ce{N2}, \ce{NH3}), or incorporated into ice mantles, influences the subsequent chemistry. These reservoirs are shaped by the physical history of the source and may ultimately determine which reaction pathways dominate. Understanding the interplay between reaction, diffusion, desorption and reactant availability is essential for understanding the observed balance between O- and N-bearing molecular reservoirs. Comparisons between chemically rich sources in different physical environments, \citep[such as G+0.693 and TMC-1 CP, where the relative contribution of O-bearing molecules differs substantially;][]{aikawa2026chemistry} support this view, suggesting that variations in molecular inventories may reflect differences in the initial reservoirs rather than a single controlling parameter.

\paragraph{\bf Modelling: }
Chemical modelling must evolve in parallel. Current models often rely on incomplete or outdated reaction networks and simplified treatments of diffusion, desorption, and destruction processes, particularly for complex and rare species. Careful attention to the chemical network is essential: users should identify the dominant reactions driving the model results and verify their reliability, ideally consulting chemists or the model developers when uncertainties arise. Discrepancies between models and observations should not be seen as failures, but as opportunities to improve our understanding of the chemistry and the network itself.
Modelling more extreme environments, such as regions with enhanced cosmic ray fluxes, strong irradiation, or varying metallicity, will help determine whether environmental effects can naturally produce the observed chemical diversity. At the same time, developers should provide clear documentation, guidelines, and maintenance of open source models to facilitate responsible use, while users should explicitly report assumptions, limitations, and sensitivity analyses in publications. This collaborative approach is essential to ensure that chemical models remain robust, interpretable, and reliable tools for understanding iCOM formation and evolution.

\paragraph{\bf Synergy: }
Finally, a \textit{strong synergy} between observers, laboratory scientists, quantum chemists, and modellers is required. Observations can guide laboratory and theoretical work by identifying key species, correlations, and non detections, while laboratory and computational results provide the fundamental parameters needed to interpret observational data. 
Advanced statistical and machine-learning approaches can further help analysing line-rich spectra, identify correlations and trends across large observational and laboratory datasets, constrain chemical networks, and quantify the relative importance of different chemical pathways. Only through this iterative approach (see Figure \ref{fig:sketch_pillars}) we can determine whether the O versus N bearing dichotomy reflects a true chemical divide, a consequence of physical conditions, or a combination of both.

\section{Conclusions} \label{sec:concl} 

This paper has examined the long-standing disparity between O- and N-bearing molecules in the interstellar medium from the complementary perspectives of observations, laboratory experiments, quantum-chemical calculations, and astrochemical modelling. 

A first conclusion is that we first need to understand how real and systematic the apparent dichotomy is. This requires consistent (i.e. in terms of facility, spatial resolution, sources and targeted species) observational surveys, which will in turn guide the next steps in laboratory experiments and quantum-mechanics computations.
The discussion of specific functional groups further illustrates this complexity. On the O-bearing side, the scarcity of molecules containing the carboxyl (\ce{-COOH}) group stands out as a key puzzle, suggesting that not all O-bearing species are equally accessible under interstellar conditions and that this chemistry is disconnected from the more common methanol-derived pathways. On the N-bearing side, the widespread presence of nitriles highlights the robustness of the \ce{-CN} group and its central role as both a reservoir of carbon and nitrogen and a likely entry point toward molecular complexity.
These trends are further shaped by environmental effects, which modulate the dominant chemical pathways and the resulting observable molecular signatures.

Secondly, we concluded that no single approach can resolve this problem. Observations reveal chemical patterns but are limited by opacity, excitation, and resolution effects. Laboratory experiments constrain reaction mechanisms, binding energies, and desorption processes under simplified conditions, while quantum-chemical calculations provide atomistic insight into reactions and surface interactions not accessible experimentally. Chemical models are required to connect these factors and produce time-dependent predictions for realistic environments. Progress therefore relies on strong integration across these approaches.

The discussions summarized in this paper also highlight several emerging well-defined questions:
whether the observed O/N differentiation is fundamentally chemical or partly observational; where and under which conditions oxygen and nitrogen reaction networks intersect; whether nitriles are mainly chemical endpoints or can be efficiently reprocessed into more saturated species; how strongly binding energies, diffusion barriers, and non-conventional surface interactions shape the chemistry; and to what extent deuteration and extragalactic environments can be used as discriminants of formation routes. In this context, sensitivity analyses have proven particularly useful in identifying the dominant parameters and reaction pathways controlling the chemical evolution.

Overall, this white paper identifies the O- versus N-bearing disparity as a framework for understanding how molecular complexity emerges, evolves in space. Its origin will only be understood through a genuinely multidisciplinary effort with broader implications for the chemistry of star- and planet-forming regions, the inheritance of interstellar material, and the pathways that may ultimately connect interstellar molecules to prebiotic chemistry.\\

\begin{acknowledgements}

We thank the referee for their valuable comments that helped improve the quality of the manuscript. We thank all participants (both on-site and in remote) of the ``Interstellar Disparity: N vs O-bearing Molecules'' 2025 Lorentz Center workshop for their active contributions, insightful discussions, and stimulating questions, which greatly shaped this white paper. 

The participants included (in alphabetical order):
Susanne Aalto,
Silvia Alessandrini,
Nadia Balucani,
Dario Barreiro-Lage,
Eleonora Bianchi,
Melisse Bonfand,
Laure Bouscasse,
Mathilde Bouvier,
Laura Busch,
Emmanuel Caux,
Cecilia Ceccarelli,
Yuan Chen,
Laura Colzi,
Marta De Simone,
François Dulieu,
Katarzyna Magdalena Dutkowska,
Joan Enrique-Romero,
Lisa Giani,
Franciele Kruczkiewicz,
Thanja Lamberts,
Ana López-Sepulcre,
Andrés Megías,
German Molpeceres,
Pooneh Nazari,
Gunnar Nyman,
Laurent Pagani,
Aina Palau,
Marten Raaphorst,
Albert Rimola,
Víctor Rivilla,
Juliette Robuschi,
Giovanni Sabatini,
Nami Sakai,
Guillaume Saury,
Riccardo Urso,
Merel van’t Hoff,
Serena Viti.\\

In particular, we deeply thank the invited speakers S. Aalto, E. Bianchi, N. Balucani, F. Dulieu, G. Molpeceres and S. Viti for providing material and constructive comments to this white paper.
We gratefully acknowledge support from the Lorentz Center and NOVA.
JER acknowledges funding from the Horizon Europe Framework Programme (HORIZON) under the Marie Skłodowska-Curie grant agreement No 101149067, ``ICE-CN''. MB acknowledges the support from the European Research Council (ERC) Advanced Grant MOPPEX 833460.\\

\end{acknowledgements}

\begin{contribution}
All the authors contributed equally to this work.
\end{contribution}

\bibliography{biblio}{}
\bibliographystyle{aasjournalv7}

\appendix
\section{Workshop information and demographics} \label{app:workshop}
\begin{figure}[ht]
    \centering
     \includegraphics[width=1\linewidth]{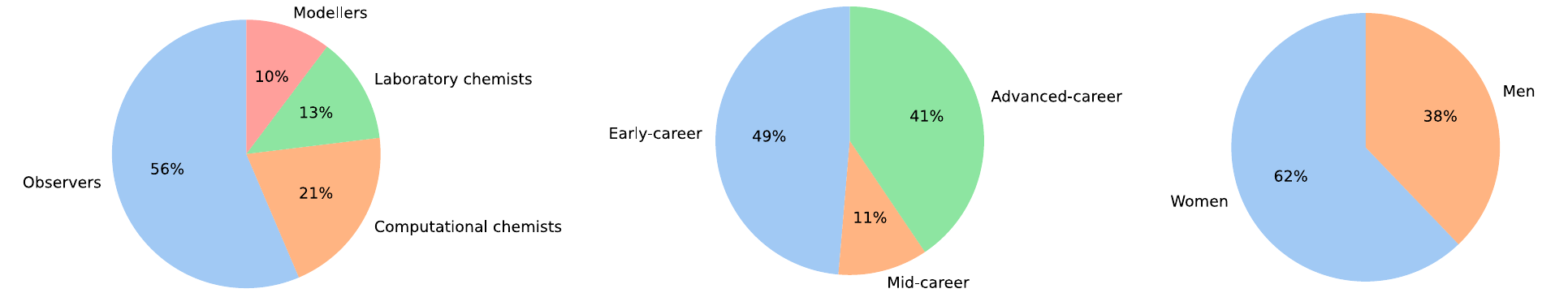}
    \caption{Demographic overview of the 37 workshop participants. The three pie charts show the distribution of participants by scientific expertise (observers, modellers, laboratory and computational chemists; \textit{left}), career stage (early, mid, and advanced career;\textit{middle}), and gender (\textit{right}).  The workshop was designed to bring together complementary expertise across observational, experimental, theoretical, and modelling approaches to interstellar chemistry.}
    \label{fig:pie-charts}
\end{figure}

\begin{figure*}[!hb]
    \centering
    \includegraphics[width=0.85\textwidth]{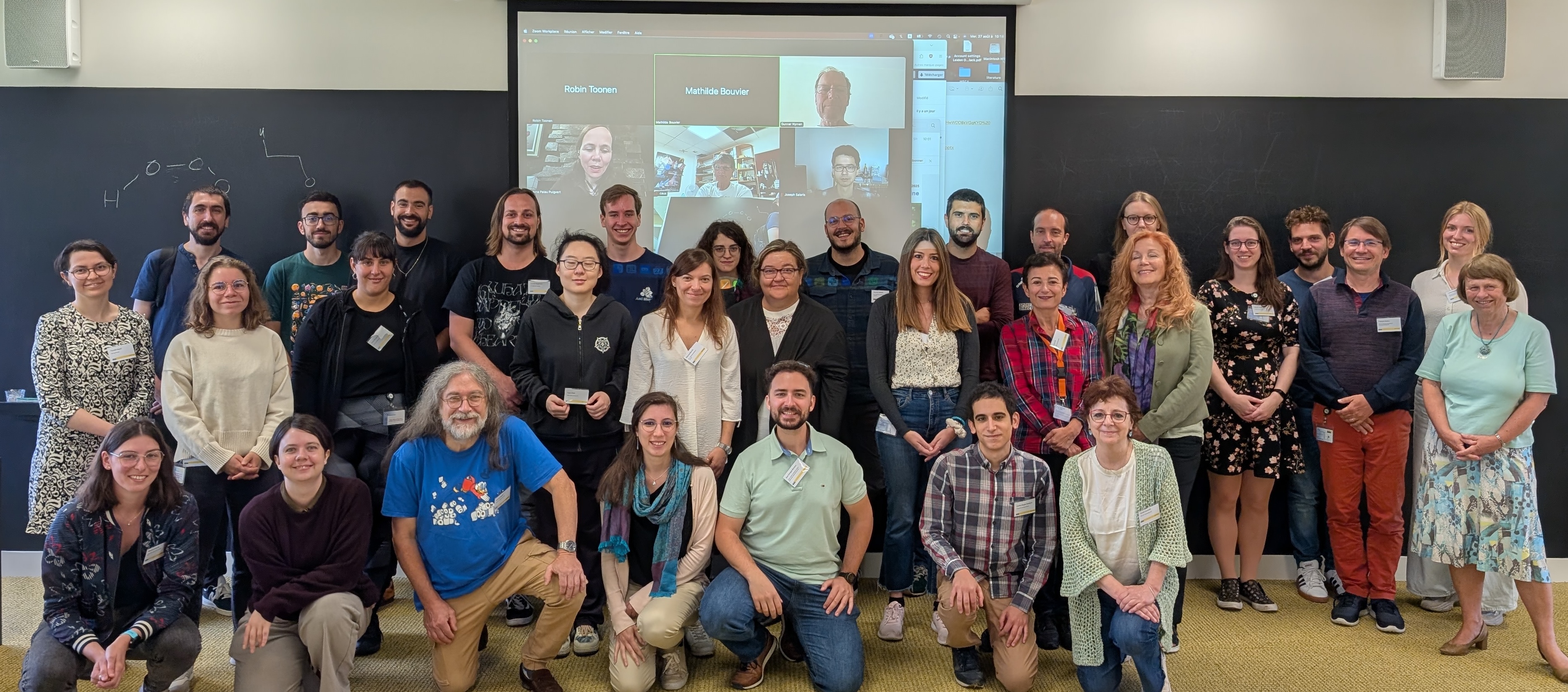}
    \caption{Participants of the Lorentz Workshop 2025 Interstellar Disparity: N vs O-bearing molecules}
    \label{fig:workshop_participants}
\end{figure*}

The results and discussions presented in this white paper are based on a Lorentz Center workshop held at the Bioscience Park of Leiden University between the 25th
and 29th of August 2025 \footnote{Workshop webpage: \url{https://www.lorentzcenter.nl/interstellar-disparity-n-vs-o-bearing-molecules.html}}.
The workshop brought together researchers from observational astronomy, laboratory astrophysics, computational chemistry, theoretical chemistry, and astrochemical modelling, with the aim of addressing the apparent differentiation between O- and N-bearing interstellar molecules. The workshop served as the basis for a community synthesis on this topic.
The workshop began each day with invited talks, followed by topic-centred breakout discussions. Participants were divided into groups to explore different themes, and the outcomes of these discussions were later brought together in a daily plenary session, where they were shared and discussed by all attendees.
A total of 37 participants attended the workshop. Figures \ref{fig:pie-charts} and \ref{fig:workshop_participants}  show the workshop participants and a summary of their demographics, including scientific expertise, career stage, and gender distribution.

\end{document}